\documentclass[conference]{IEEEtran}

\usepackage{cite}
\usepackage{color}
\usepackage{multirow}
\usepackage{booktabs,multirow,array}
\usepackage[table]{xcolor} 
\usepackage{ltxcmds}
\usepackage{footnote}
\usepackage{amsthm}

\usepackage[table]{xcolor}
\usepackage{subfigure}

\usepackage{balance}

\usepackage{lipsum}
\usepackage{amssymb} 
\usepackage{pdflscape}
\usepackage{afterpage}
\usepackage{tabularx}

\ifCLASSINFOpdf
\usepackage[pdftex]{graphicx}
\else
  \usepackage[dvips]{graphicx}
\fi

\begin{document}
\title{Security of Machine Learning-Based Anomaly Detection in Cyber Physical Systems}


\author{\IEEEauthorblockN{Zahra Jadidi$^{\ast}$, 
Shantanu Pal$^{\ast}$, Nithesh Nayak K$^{\ast}$, Arawinkumaar Selvakkumar$^{\ast}$, Chih-Chia Chang$^{\ast}$, \\ Maedeh Beheshti$^{\mathsection\mathsection}$, Alireza Jolfaei$^{\ast\ast}$}
\IEEEauthorblockA{$^{\ast}$School of Computer Science, Queensland University of Technology, Brisbane, QLD 4000, Australia\\ 
{$^{\mathsection\mathsection}$Critical Path Institute, Tucson, AZ 85718, USA}\\
{$^{\ast\ast}$School of Computer Science, Macquarie University, Sydney, NSW 2109, Australia}\\
{zahra.jadidi@qut.edu.au,
shantanu.pal@qut.edu.au,
nithesh.nayakkinnimulky@connect.edu.au, 
mbeheshti@c-path.org,}\\
{chihchia.chang@connect.qut.edu.au, 
arawinkumaar.selvakkumar@connect.qut.edu.au,
alireza.jolfaei@mq.edu.au} {}}}




\maketitle


\begin{abstract}
With the emergence of the Internet of Things (IoT) and Artificial Intelligence (AI) services and applications in the Cyber Physical Systems (CPS), the methods of protecting CPS against cyber threats is becoming more and more challenging. Various security solutions are implemented to protect CPS networks from cyber attacks. For instance, Machine Learning (ML) methods have been deployed to automate the process of anomaly detection in CPS environments. The core of ML is deep learning. However, it has been found that deep learning is vulnerable to adversarial attacks. Attackers can launch the attack by applying perturbations to input samples to mislead the model, which results in incorrect predictions and low accuracy. For example, the Fast Gradient Sign Method (FGSM) is a white-box attack that calculates gradient descent oppositely to maximize the loss and generates perturbations by adding the gradient to unpolluted data. In this study, we focus on the impact of adversarial attacks on deep learning-based anomaly detection in CPS networks and implement a mitigation approach against the attack by retraining models using adversarial samples. We use the Bot-IoT and Modbus IoT datasets to represent the two CPS networks. We train deep learning models and generate adversarial samples using these datasets. These datasets are captured from IoT and Industrial IoT (IIoT) networks. They both provide samples of normal and attack activities. The deep learning model trained with these datasets showed high accuracy in detecting attacks. An Artificial Neural Network (ANN) is adopted with one input layer, four intermediate layers, and one output layer. The output layer has two nodes representing the binary classification results. To generate adversarial samples for the experiment, we used a function called the `fast\_gradient\_method' from the Cleverhans library. The experimental result demonstrates the influence of FGSM adversarial samples on the accuracy of the predictions and proves the effectiveness of using the retrained model to defend against adversarial attacks.

\end{abstract}

\begin{IEEEkeywords}
Cyber physical systems, Machine learning, Security, Attacks, Defence, Internet of Things.
\end{IEEEkeywords}

\IEEEpeerreviewmaketitle

\section{Introduction}
\label{introduction}

Cyber Physical Systems (CPS) are interconnected computer systems that connect physical devices to the cyber world to operate a process efficiently \cite{pivoto2021cyber}. 
The emergence of the Internet of Things (IoT) in CPS plays a key role in the fourth industrial revolution (i.e., Industry 4.0) \cite{yaacoub2020cyber}. There has been a significant trend to deploy IoT applications and services to monitor and control CPS \cite{sharma2022role}. 
Alongside the exponential increase in the number of IoT devices in CPS, networks is the growth of unknown vulnerabilities and threats posing a big security challenge in the present preventive systems. This challenge in the CPS network increases the surface of attacks allowing adversaries to gain control of unsecured devices~\cite{pal2017design} \cite{pal2021analysis}. Towards this, the Intrusion Detection Systems (IDS) provides successful ways to classify and identifying attacks in CPS networks. Traditional IDS use different methods to identify and detect different attacks, for example, through statistical-based and patterned detection. New features are also applied in IDS enhancing their detection capabilities through behavioural patterns without relying on built-in signatures. Machine Learning (ML) techniques, on the other hand, are integrated into IDS to support automated decision making in the classification and identification of any greater array of attacks. In particular, a recent study \cite{da2019internet} shows that the majority of the state-of-the-art IDS are built on systems, e.g.,  Support Vector Machines (SVM), Random Forest, and Decision Trees. However, other studies \cite{al2020survey} \cite{haylett2021system} also show that deep learning-based IDS has more advantages over conventional IDS because of its performance in large datasets and automated extraction of complex representation from data.

In regard to this context, the enormous usage of different types of ML and deep learning 
models have attracted adversarial entities by implementing noise and disrupting prediction outcomes that cause a result to be invalid or incorrect. This is known as `Adversarial Machine Learning' (AML), which can exploit the weakness of a pre-trained model by manipulating data and network traffic that traverse through IDS devices \cite{anthi2021hardening}. The perturbations engaged by the adversary in input data can reduce the model's effectiveness in its deciding boundaries wherein malicious packets are misclassified as benign. Following this notion, gradient-based adversarial attacks \cite{ozbulak2020perturbation} are proven to be highly effective in adding interference in gradient inputs of the model. Examples of these adversarial attacks are the Fast Gradient Sign Method (FGSM) \cite{goodfellow2014explaining}, the Jacobian-based Saliency Map Attack (JSMA) \cite{papernot2016limitations}, Deepfool \cite{moosavi2016deepfool}, and Carlini Wagner attack (CW) \cite{carlini2017towards}. To this end, to enhance the capability of a deep learning-based IDS, different models, approaches, and techniques are used to determine its effectiveness in preventing such adversarial attacks from disrupting its decision making. 

An Artificial Neural Network (ANN) is one of the well-established methods to represent the classification of either numeric or image data \cite{abiodun2018state}. 
ANNs have been deployed by many papers to perform anomaly detection in CPS networks \cite{wu2020anomaly} \cite{luo2021deep}. However, deep learning-based anomaly detection methods are vulnerable to AML attacks. 
In accordance with the vulnerability of the deep learning methods, there is a huge demand for finding defensive techniques to improve robustness against adversarial attacks. However, in the existing studies, the results still show a significant gap in addressing the gradient-based adversarial attacks.

To address this issue, in this paper, we investigate the impact of FGSM attacks on deep learning-based anomaly detection in CPS networks and present a defensive strategy for these attacks. 
To the best of our knowledge, unlike other studies in this domain, we intend to investigate FGSM-based attacks in CPS networks including Industrial IoT systems. The major contributions of the paper can be summarized  as follows:

\begin{itemize}
    \item We explore the effects of adversarial attacks on deep learning-based anomaly detection in a CPS network using real-world Industrial IoT datasets.
    
    \item Based on our findings, we design a framework to address the impact of FGSM adversarial attacks on deep learning-based anomaly detection in CPS networks.  
    
    \item We implement the framework to mitigate the  attacks  by retraining our novel neural network-based solution using adversarial samples.
    
    \item We evaluated the model results using our generated adversarial attacks with variation in noise. Our results show how the model robustness improved 92.8\% after using the proposed defence strategy.

    
\end{itemize}

The rest of the paper is organized as follows: In Section~\ref{Review of Related Literature}, we discuss related work. In Section~\ref{Methodology}, we present a detailed methodology used in our design. In Section~\ref{Results and Discussion}, we show the results achieved from the experiments and provide a discussion of the significant findings. Section~\ref{Defence strategies} discusses the defence strategies. Finally, in Section~\ref{conclusion}, we conclude the paper with future work.

\section{Related Work}
\label{Review of Related Literature}

Several studies have been conducted to apprehend and mitigate the effects of cyber attacks on CPS using ML \cite{kim2021survey,mujeeb2021machine,jadidi2020securing}. 
Nevertheless, there has been less emphasis on AML attacks to ML-based solutions. 
The vulnerabilities of ML-based solutions have been reviewed in various areas, e.g., 
spam detection, malware detection, and intrusion detection in computer networks. There are different types of attacks that can exploit these vulnerabilities, e.g. FGSM and 
Generative Adversarial Networks (GANs).

Proposal \cite{zhang2021label} demonstrates a type of poisoning attack called label flipping that can reduce a performance metrics of a classification model by flipping labels on the sample data. In addition, proposal \cite{patil2021improving} evaluates the robustness of different classification models by implementing a malware classification system through ML. The results of this experiment show that the models, when loaded with FGSM attack, can gravely affect the performance results, and when adversarial training is executed, the model becomes robust with the detection. 
In \cite{chauhan2020polymorphic}, the authors demonstrate a study by using the GAN model to generate adversarial Distributed Denial of Service (DDoS) attacks 
that are successful in evading IDS because of their polymorphic ability to change signatures. The papers discussed above studied AML attacks in typical IT networks, and they did not work on CPS networks with industrial devices which follow different behavioural patterns. 
In our study, we focus specifically on the CPS networks.

In the context of CPS, there is not sufficient research on AML attacks on ML-based IDS. 
AML attacks in CPS networks have been reviewed by a number of papers \cite{song2020fda, dankwa2021securing,qiu2020adversarial}. 
In health systems immunity and vulnerability, while integrated with deep learning algorithms, proposal \cite{mode2020crafting} investigates different adversarial attacks like FGSM and Basic Iterative Method (BIM) that show the effects of the amount of perturbation versus the performance variation.
An encode-decode model has been proposed by \cite{yang2021iot} to provide evidence of the robustness against FGSM only on a limited number of IoT datasets. 
The authors of \cite{luo2020adversarial} 
propose an AML-based partial-model attack. This attack could be performed by changing a small part of the device data. They show that ML-based analysis in IoT systems are vulnerable to AML attacks even if the adversaries manipulate a small portion of the device data. 

The authors of \cite{ibitoye2019analyzing} use the Bot-IoT dataset to implement several adversarial attacks, namely FGSM, BIM and Projected Gradient Descent (PGD), against two deep learning methods Forward-Feed Neural Network (FNN) and Self-Normalizing Neural Network (SNN), to compare the results of various gradient-based adversarial attacks. The authors demonstrate that adversarial samples have a different impact on both deep learning models and show that SNN has more resiliency against adversarial samples. Additionally, the demonstrations show that when dataset features are normalized, performance metrics are better but are more susceptible to adversarial attacks. However, the defence strategy against these attacks was not proposed in this paper. Proposal \cite{wang2018deep} also implements a deep learning algorithm-based IDS that uses multi-layer perceptron FNN against different gradient-based attacks, and the metrics show that the FNN had been gravely affected by the initiated adversarial attacks. 

The aforementioned proposals did not consider AML attacks in Industrial IoT which are known as important components of CPS. In this study, we will investigate the impact of AML attacks on both IoT and Industrial IoT. In addition, unlike these studies, in this paper, we focus on 
creating different variants of gradient-based attacks with the help of ML and use that instance in hardening the defence of a classification model continuously by implementing an iterative retraining approach. 
In summary, unlike the above-mentioned proposals, we focus on implementing an architecture that can self-sustain its defences against adversarial attacks in CPS at scale. Additionally, for experimental purposes, the datasets that have been used in most of the previous studies do not include Industrial IoT in CPS networks. They also do not investigate the required defence strategy for Industrial IoT in the context of CPS networks, which we consider in this paper. Further, in this paper, we particularly address this gap by proposing a defensive strategy against FGSM attacks in IoT and Industrial IoT networks.

\section{Methodology}
\label{Methodology}
\begin{figure}[ht]
    \centering
    \includegraphics[scale=.75]{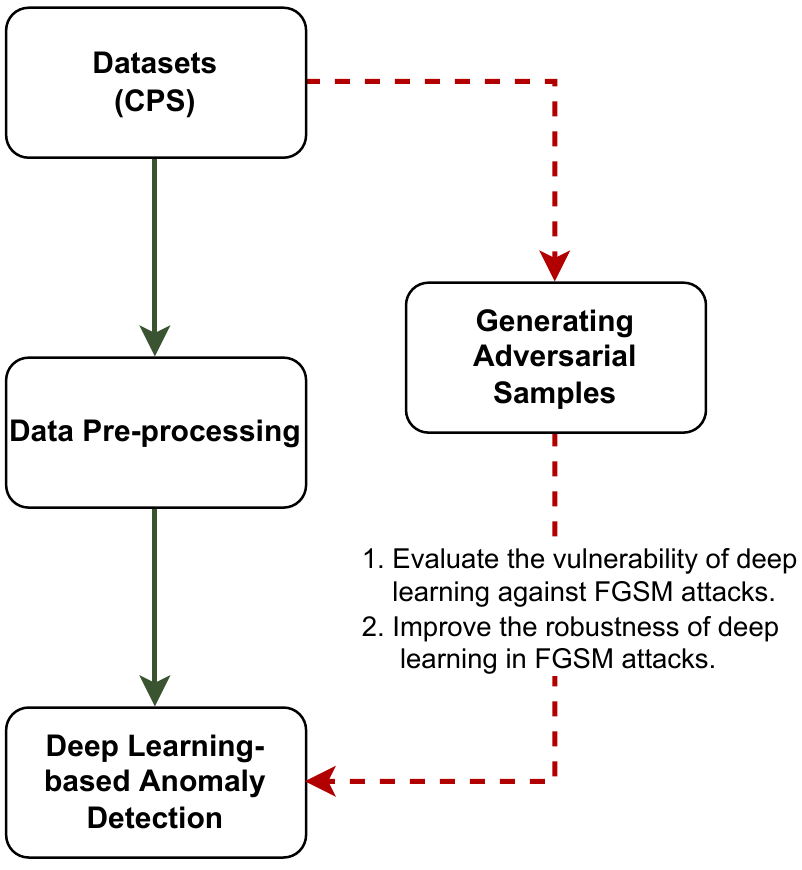}
    \caption{The functional components of the proposed model.}
    \label{arch}
\end{figure}

In this section, we detail the methodology. Before discussing it in detail, we present how our proposed model would function. In Fig.~\ref{arch}, we illustrate the functional components of the proposed model. Initially, we discuss the datasets (i.e., IoT and Industrial IoT datasets) used in this study to investigate the impact of FGSM attacks on deep learning-based security solutions. Then, the deep learning-based anomaly detection and FGSM model will also be explained. 


\subsection{Datasets}
\label{Dataset}
In our experiment, we have used two datasets, namely \textit{Bot-IoT} and \textit{Modbus}. These datasets are provided by the Cyber Range Lab of UNSW Canberra, Australia. A brief introduction of each of them are as follows~\cite{dataset-link}:

\subsubsection{Bot-IoT Dataset}  This dataset 
was comprised of over 72-million IoT network traffic records originally \cite{koroniotis2019towards}. In order to make the data processing more efficient, 5\% of the dataset was extracted resulting in approximately 3.6 million records \cite{koroniotis2019towards}. The dataset has 19 features, and in this study, we selected the 10 best features as shown in Table \ref{table:10_best_features}. The selection was based on the Correlation Coefficient and Joint Entropy Score discussed in \cite{koroniotis2019towards}. Five classes are contained in the dataset including (i) DDoS, (ii) DoS, (iii) Reconnaissance, (iv) Theft and (v) Normal. All the classes except Normal were labelled as “1”, which stands for attack, while normal records were labelled as “0”.

\begin{table}[h]
    \caption{Selection of the ten best features from Bot-IoT dataset \cite{koroniotis2019towards}.}
    \small
    \scalebox{.92}{
    \begin{tabular}{l l}
        \toprule
            Features & Description \\
        \toprule
            seq & Argus sequence number\\
            stddev & Standard deviation of aggregated records\\
            N\_IN\_Conn\_P\_SrcIP & Total number of packets per source IP \\
            min & Minimum duration of aggregated records \\
            state\_number & Numerical representation of transaction state\\
            mean & Average duration of aggregated records\\
            N\_IN\_Conn\_P\_DstIP & Total number of packet per destination IP \\
            drate & Destination-to-source packets per second \\
            srate & Source-to-destination packets per second \\
            max & Maximum duration of aggregated records \\
        \bottomrule
    \end{tabular}
    }
    \label{table:10_best_features}
\end{table}

\subsubsection{Modbus Dataset} 
This is a benchmark dataset  for Industry 4.0/IoT and Industrial IoT (IIoT) for evaluating the fidelity and efficiency of various cybersecurity applications based on Artificial Intelligence (AI). It comprised   around 50 thousand records. It is one of the datasets from the TON\_IoT Telemeter dataset \cite{alsaedi2020ton_iot}, and contains records captured by Modbus sensors. This dataset is composed of 8 features as shown in Table \ref{table:modbus_features} and categorized into four classes containing (i) DoS, (ii) DDoS, (iii) Backdoor attack and (iv) Normal. 



\begin{table}[h]
    \caption{IoT Modbus dataset features and description \cite{alsaedi2020ton_iot}.}
    \small
    \scalebox{.92}{
    \begin{tabular}{p{3.5cm} p{5cm}}
    \toprule
    Features & Description \\
    \toprule
        ts & Timestamp of sensor reading data \\
        date & Date of logging Modbus register's data \\
        time & Time of logging Modbus register's data \\
        FC1\_Read\_Input\_Register & Modbus function code that is responsible for reading an input register \\
        FC2\_Read\_Discrete\_Value & Modbus function code that is in charge of reading a discrete value \\
        FC3\_Read\_Holding\_Register & Modbus function code that is responsible for reading a holding register \\
        FC1\_Read\_Coil & Modbus function code that is responsible for reading a coil \\
        type & A tag with normal or attack subclasses, e.g., DoS, DDoS, and backdoor attacks\\
        \bottomrule
    \end{tabular}
    }
    \label{table:modbus_features}
\end{table}

\subsection{Data Pre-processing}
\label{Imbalanced_Data}

The original Bot-IoT dataset was comprised of unhandled data which this experiment needs to address, as shown in Table~\ref{tab:my_label2}. Redundant and irrelevant features in the data can cause a problem in the classification model in predicting outcomes, especially when dealing with large dimension datasets~\cite{abdulwahab2022feature}. To appropriately conduct this experiment, the Bot-IoT dataset goes through different data pre-processing techniques that help identify relevant data from insignificant information. As the first step of the data pre-processing, the dataset has to be standardized using functions that will remove redundant and unused columns and rows in the dataset. Secondly, the dataset undergoes the process of Synthetic Minority Oversampling Technique (SMOTE)~\cite{pears2014synthetic} that uses random samples to balance the distribution of the output labels and  addresses the effects of imbalanced data, as shown in Table ~\ref{tab:my_label2}. Unlike the Bot-IoT dataset, the Modbus dataset has class distribution close to each other, and there was no significant difference in results after balancing, due to which we do not apply any data balancing technique for the Modbus dataset.

Finally, once the Bot-IoT dataset distribution is balanced, it is prepared to convert categorical values into numerical data, since Modbus dataset is normalized without the need of data balancing. Once the dataset is normalized, it is split into training and testing datasets in our classification model, as shown in Table~\ref{table:train_test_amount}.


\begin{table}[h]
    \centering
    \caption{Imbalanced and balanced datasets.}
    \small
    \scalebox{.85}{
    \begin{tabular}{l l l l}
    \toprule
        & & Imbalanced Dataset & Balanced Dataset \\
    \toprule
        Dataset Name& Output Labels & Total Samples & Total Samples \\
    \hline
        Bot-IoT & Attack & 293,447 & 293,447 \\
        & Normal & 370 & 293,447 \\
    \hline
        Modbus & Attack & 13,886 & ...\\
        & Normal & 17,476 & ...\\
    \bottomrule
    \end{tabular}
    }
    \label{tab:my_label2}
\end{table}

\begin{table}[h]
    \centering
    \caption{Training and testing data amount.}
    \small
    \scalebox{.95}{
    \begin{tabular}{l l l l}
    \toprule
        & Total & Training Samples & Testing Samples \\
        & & 70\% & 30\% \\
    \toprule
        Bot-IoT & & & \\
        (imbalanced) & 2,934,817 & 2,054,371 & 880,446 \\
    \hline
        Bot-IoT & & \\
        (balanced) & 5,868,894 & 4,108,225 & 1,760,669 \\
    \hline
        Modbus & 51,106 & 21,953 & 9,409 \\
    \bottomrule
    \end{tabular}
    \label{table:train_test_amount}
    }
\end{table}

\subsection{Deep Learning-Based Anomaly Detection}
\label{Deep-learning}
In this study, the classification model used was ANN networks obtained from \cite{papadopoulos2021launching}. 
ANN is composed of one input layer, four hidden layers and one output layer. The underlying model architectures are the same, due to variation in features of each dataset the input layer is different (Bot-IoT dataset has 10 nodes, while the Modbus dataset has 8 nodes). 
The hidden layers are activated with TanH and they are composed of 20, 60, 80 and 90 nodes, respectively. Both output layers use Sigmoid to activate and generate binary results (1 for attack or 0 for normal).




\subsection{Generating Adversarial Samples}
\label{Generative Adversarial Samples}
In recent studies performed regarding adversarial attacks on ML, gradient-based attacks pose a severe threat to neural networks \cite{mahfuz2021mitigating}. Thus, a common gradient-based attack was employed to demonstrate how badly it can affect the classification model's accuracy and use that instance to create more devastating adversarial attacks to be used in the proposed approach.
Recall, we use FGSM for generating adversarial examples \cite{goodfellow2014explaining}. FGSM is performed based on Eq.~\ref{equ:fgsm}, which maximize the loss by calculating the gradient descent oppositely. In Eq.~\ref{equ:fgsm}, $\widetilde{x}$ stands for adversarial samples, $x$ represents the input data, and $J$ is the loss function. As shown in the function, $\epsilon$ is a crucial variable that can significantly affect the output. The $\epsilon$ value is used to add noise which leads to worse performance.

\begin{equation}
\label{equ:fgsm}
    \widetilde{x} = x + \epsilon \cdot sign(\nabla_{x}J(w,x,y))
\end{equation}

In order to generate adversarial samples, the function `\texttt{fast\_gradient\_method}' from CleverHans v2.1.0 \cite{papernot2016technical} is used, which is a library for adversarial examples. As the function was written following the equation mentioned above, all the values of variables in the equation can be changed via the parameters of the function. 


\section{Results and Discussion}
\label{Results and Discussion}
This section will discuss the evaluation metrics employed for the experiment. We also present the results and discuss the significant findings.
\subsection{Evaluation Metrics}
\label{Evaluation Metrics}
Evaluating specific metrics must be well defined to find out how well the proposed approach is. The study used the evaluation metrics indicators that assess the experimental results, i.e., the accuracy, precision, recall, and F1-score, Eqs.~\ref{eqn:eq1},~\ref{eqn:eq2}, and ~\ref{eqn:eq3}. where \textit{(TP)} is true positive, \textit{(TN)} is true negative, \textit{(FP)} is false positive, and \textit{(FN)} shows false negative.

\begin{equation}
\label{eqn:eq1}
    Precision=\frac{TP}{TP+FP}
\end{equation}

\begin{equation}
\label{eqn:eq2}
    Recall=\frac{TP}{TP+FN}
\end{equation}

\begin{equation}
\label{eqn:eq3}
    F1\_Score=\frac{2(Precision \cdot Recall)}{Precision+Recall}
\end{equation}

These evaluation metrics are summarized succinctly using the confusion matrix 
which evaluates model proficiency with a prediction for the class.



\subsection{Results}
\label{Results}

\begin{figure}[ht] 
    \centering
    \includegraphics[scale=.32]{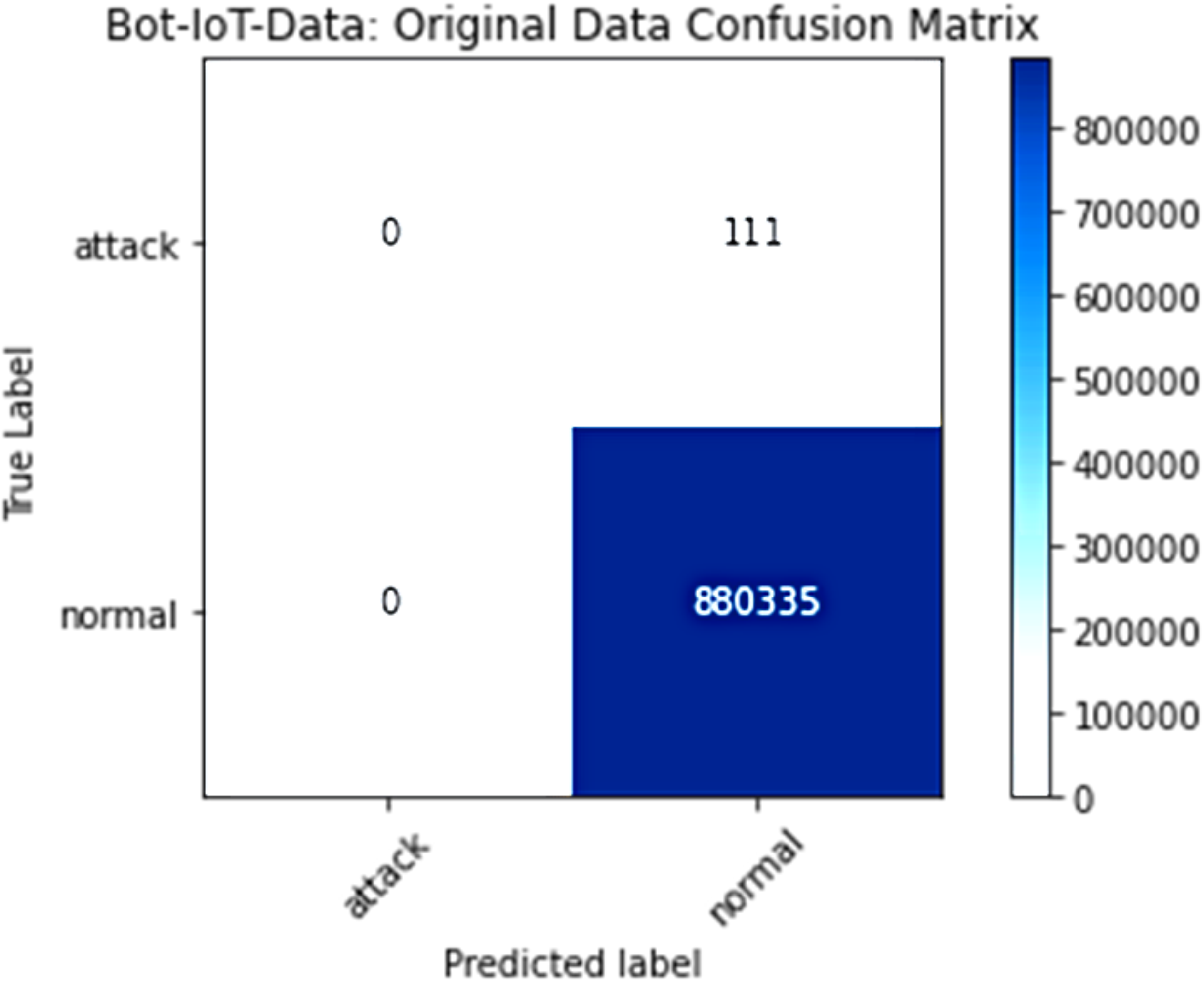}
    \caption{Confusion matrix in imbalanced Bot-IoT dataset.}
    \label{fig-unbalanced_original}
\end{figure}

\begin{figure}[ht]  
    \centering
    \includegraphics[scale=.34]{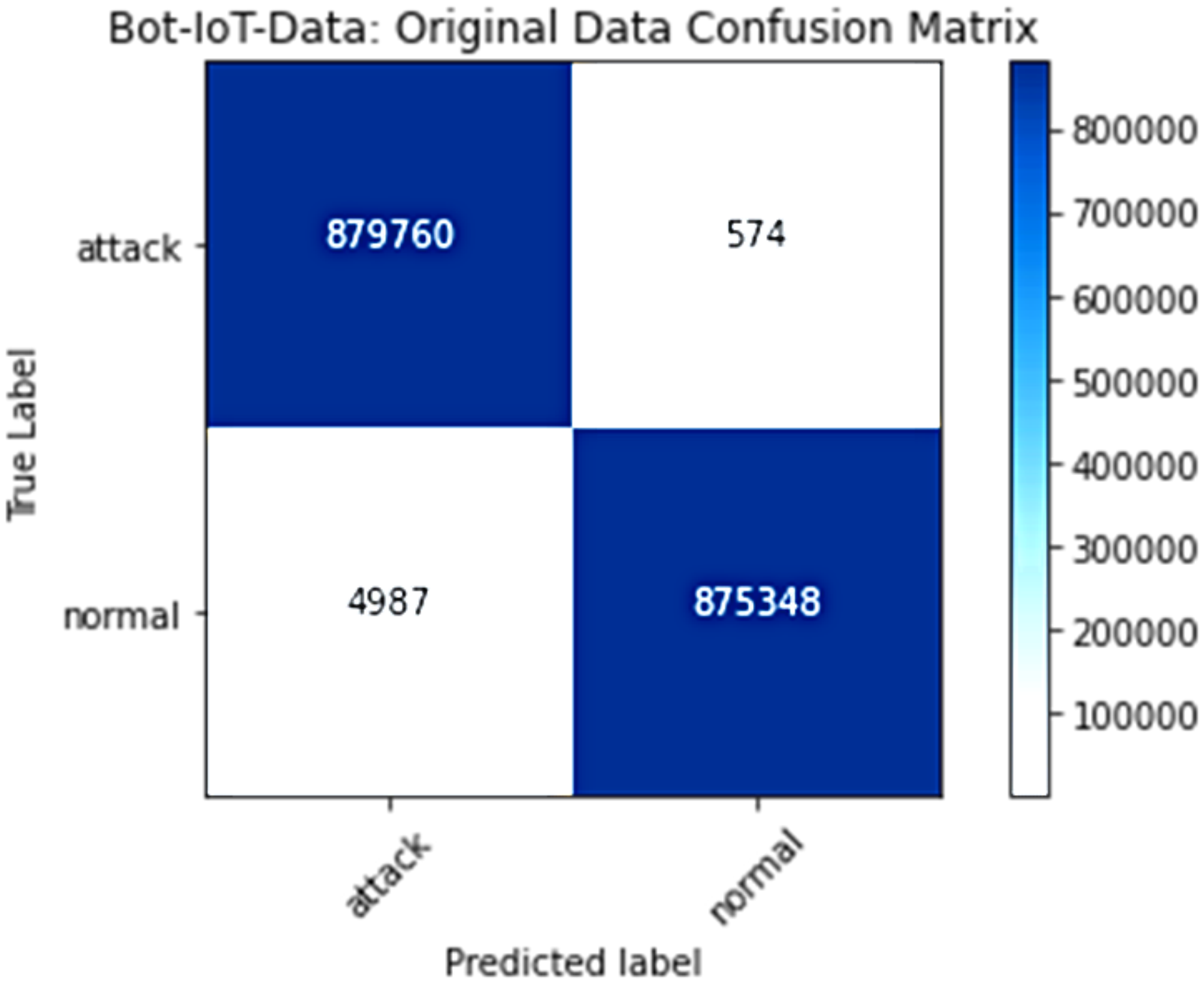}
    \caption{Confusion matrix of ANN in balanced Bot-IoT.}
    \label{fig:balanced-bot}
\end{figure}

\begin{figure}[ht]   
    \centering
    \includegraphics[scale=.32]{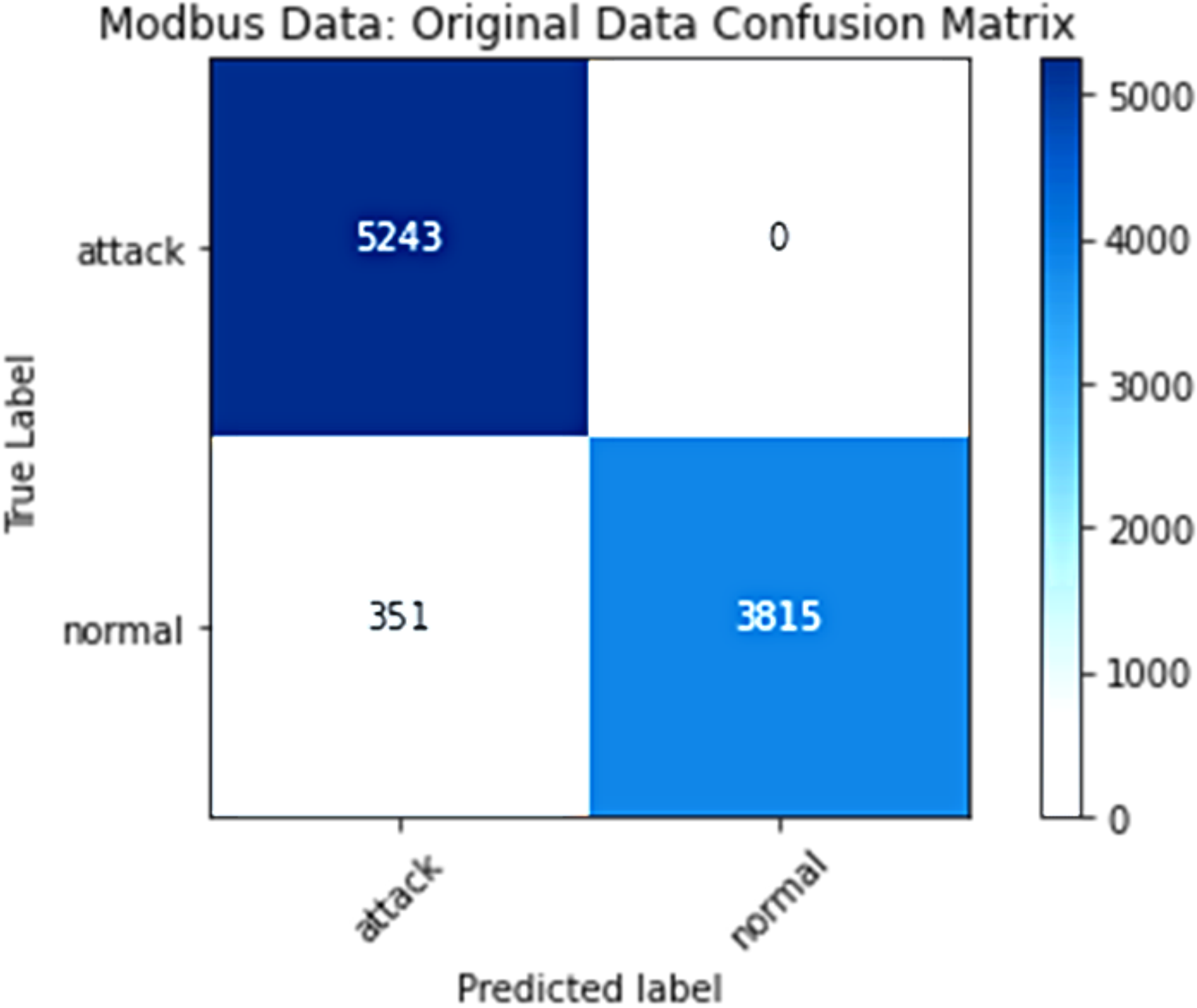}
    \caption{Confusion matrix of ANN in Modbus dataset.}
    \label{fig:modbus-original}
\end{figure}

In this section, we present the performance of the ANN classifier under FGSM attacks for both Bot-IoT and Modbus datasets. Also, experiments have been conducted for the Bot-IoT dataset for scenarios handling imbalanced datasets. The functional codes of our solution can be
found in the following repository \cite{git-code}. The results show how the imbalanced data resulted in an over-fitting model. We have 
trained the model for three scenarios, two with the Bot-IoT dataset (balanced and unbalanced), and another one with the Modbus dataset. The results of these three data scenarios can be seen in Fig.~\ref{fig-unbalanced_original}, Fig.~\ref{fig:balanced-bot} and Fig.~\ref {fig:modbus-original}. We can also observe from Fig.~\ref{fig-unbalanced_original} and Fig.~\ref{fig:balanced-bot} the improvements between the imbalanced dataset and balanced dataset, which manifest the need for the data balance in the Bot-IoT dataset.

\begin{table}[h]
    \centering
    \caption{Evaluation results in Bot-IoT and Modbus datasets.}
    \small
    \scalebox{.78}{
    \begin{tabular}{l l l l l l}
    \toprule
       Datasets & Testing Dataset & TP & TN & FP & FN \\
    \toprule
  Bot-IoT dataset   & Original dataset	& 0	& 880,335	& 0	& 111\\
(imbalanced)	& FGSM epsilon=0	&0	 &880,335 &0	&111\\
	& FGSM epsilon=0.2	&41	 &879,400 &935 &70\\
	&FGSM epsilon=0.4	&98	 &852,893 &27,442 &13\\
	&FGSM epsilon=0.6	&111 &712,199 &168,136 &0\\
	&FGSM epsilon=0.8	&111 &412,583 &467,752 &0\\
	&FGSM epsilon=1	    &111 &63,022  &817,313	&0\\
    \hline
  Bot-IoT dataset  & Original dataset & 879,760 & 875,348 & 4,987 & 574\\
 (balanced)	&FGSM epsilon=0	  & 879,760	& 875,348 & 4,987   & 574\\
	&FGSM epsilon=0.2 & 577	    & 802,486 & 77,849  & 879,757\\
	&FGSM epsilon=0.4 & 577	    & 397,280 & 483,055 & 879,757\\
	&FGSM epsilon=0.6 & 577	    & 35,262  & 845,073 & 879,757\\
	&FGSM epsilon=0.8 & 577	    & 4976	  & 875,359 & 879,757\\
	&FGSM epsilon=1	  & 577	    & 4976	  & 875,359 & 879,757\\
    \hline
    Modbus dataset	&Original dataset	&5,243	&3,815	&351	&0\\
	&FGSM epsilon=0	    &5,243	&3,815	&351	&0\\
	&FGSM epsilon=0.2	&4,796	&2,222	&1,944	&447\\
	&FGSM epsilon=0.4	&806	&218	&3,948	&4,437\\
	&FGSM epsilon=0.6	&0	    &203	&3,963	&5,243\\
	&FGSM epsilon=0.8	&0	    &203	&3,963	&5,243\\
	&FGSM epsilon=1	&0	&203	&3,963	&5,243\\
    \bottomrule
    \end{tabular}
    }
    \label{tab:results}
\end{table}

\begin{table}[h]
    \centering
    \caption{Evaluation matrics in both Bot-IoT and Modbus datasets.}
    \small
    \scalebox{.86}{
    \begin{tabular}{l l l l l }
    \toprule
       Datasets & Testing dataset & Precision  & Recall & F1-Score \\
       {} &  & (\%) & (\%)& \\
    \toprule
Bot-IoT dataset     & Original dataset	& 0 & 0& 0\\
(imbalanced)	& FGSM epsilon=0	&0	&0&0\\
	& FGSM epsilon=0.2	&4.2&36.94&7.54\\
	&FGSM epsilon=0.4	&0.36&88.29&0.71\\
	&FGSM epsilon=0.6	&0.07&100&0.13\\
	&FGSM epsilon=0.8	&0.02&100&0.05\\
	&FGSM epsilon=1	    &0.01&100&0.03\\
    \hline
  Bot-IoT dataset   & Original dataset	&99.44&99.93&99.68\\
(balanced)	&FGSM epsilon=0	   &99.44&99.93&99.68\\
	&FGSM epsilon=0.2	&0.74&0.07&0.12\\
	&FGSM epsilon=0.4	&0.12&0.07&0.08\\
	&FGSM epsilon=0.6	&0.07&0.07&0.07\\
	&FGSM epsilon=0.8	&0.07&0.07&0.07\\
	&FGSM epsilon=1	    &0.07&0.07&0.07\\
    \hline
    Modbus dataset	&Original dataset	&93.73&100&96.76\\
	&FGSM epsilon=0	    &93.73&100&96.76\\
	&FGSM epsilon=0.2	&71.16&91.47&80.05\\
	&FGSM epsilon=0.4	&16.95&15.37&16.12\\
	&FGSM epsilon=0.6	&0&0&0\\
	&FGSM epsilon=0.8	&0&0&0\\
	&FGSM epsilon=1	    &0&0&0\\
    \bottomrule
    \end{tabular}
    }
    \label{tab:matrics}
\end{table}

\begin{figure}[ht]  
    \centering
    \includegraphics[scale=.35]{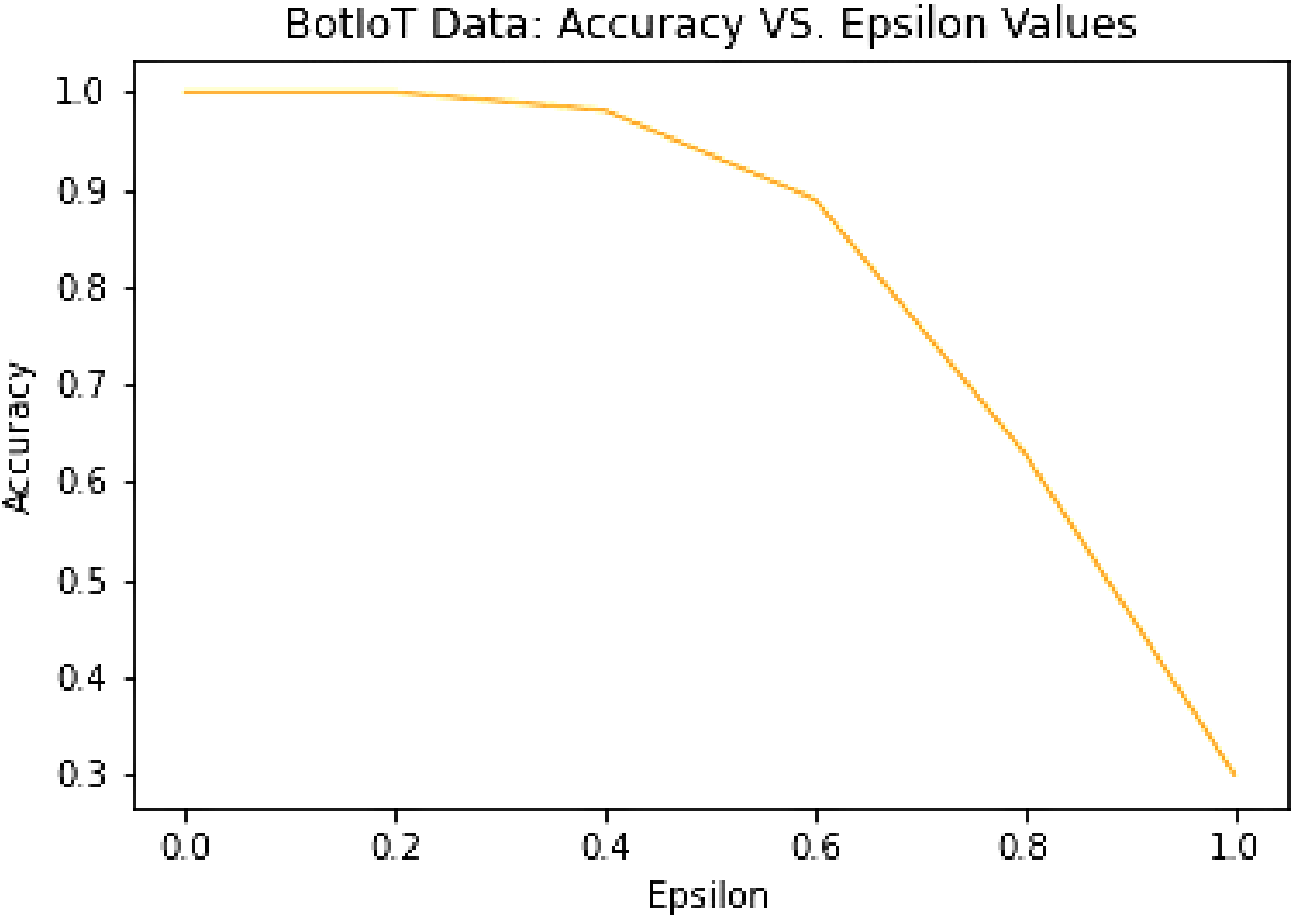}
    \caption{Impact of Epsilon values on the prediction accuracy (for imbalanced Bot-IoT dataset).}
    \label{fig:epsilon-imbalanced-bot}
\end{figure}

\begin{figure}[ht]   
    \centering
    \includegraphics[scale=.34]{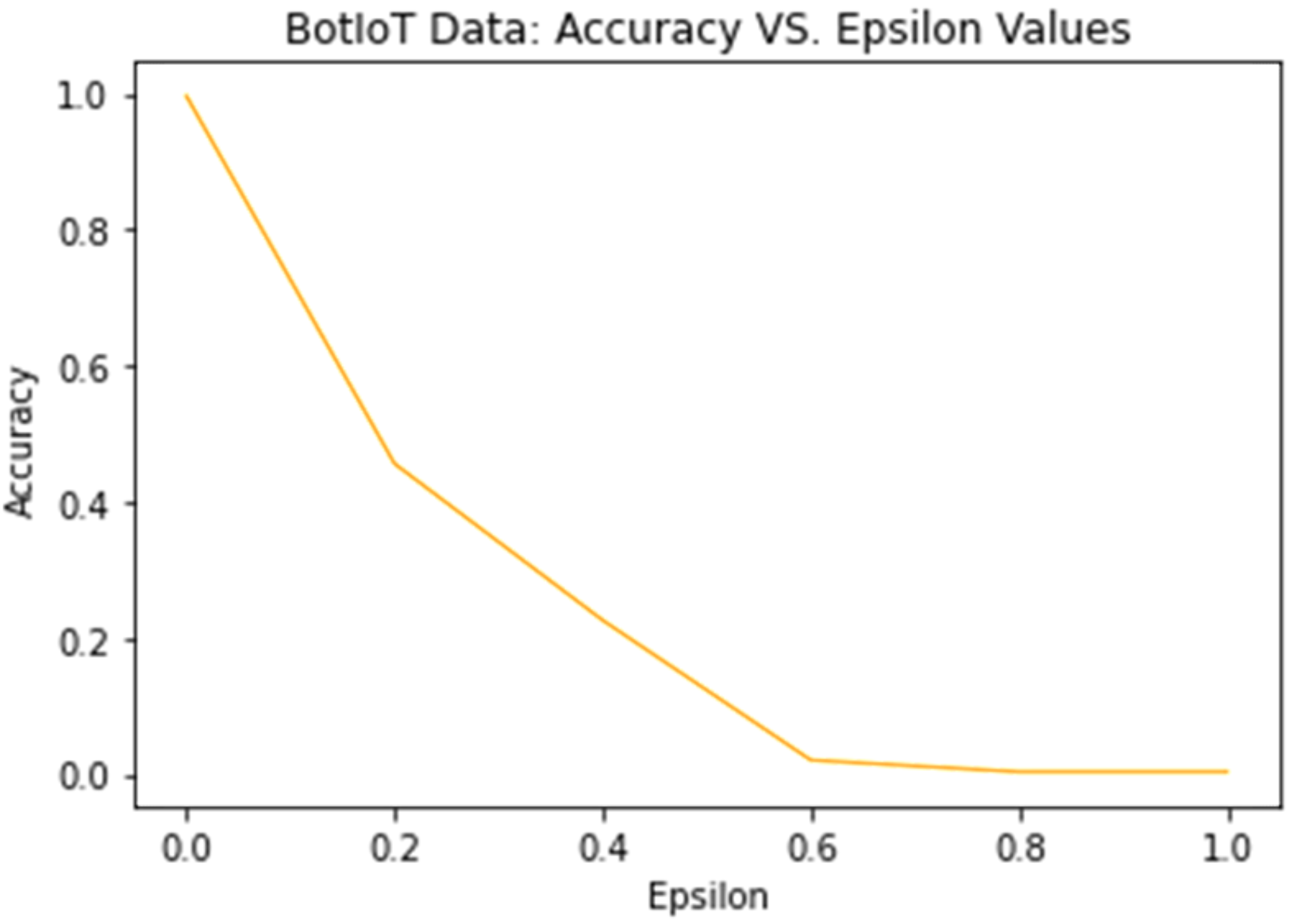}
    \caption{Impact of Epsilon values on the prediction accuracy (for balanced Bot-IoT dataset).}
    \label{fig:epsilon-balanced-bot}
\end{figure}

\begin{figure}[ht]   
    \centering
    \includegraphics[scale=.35]{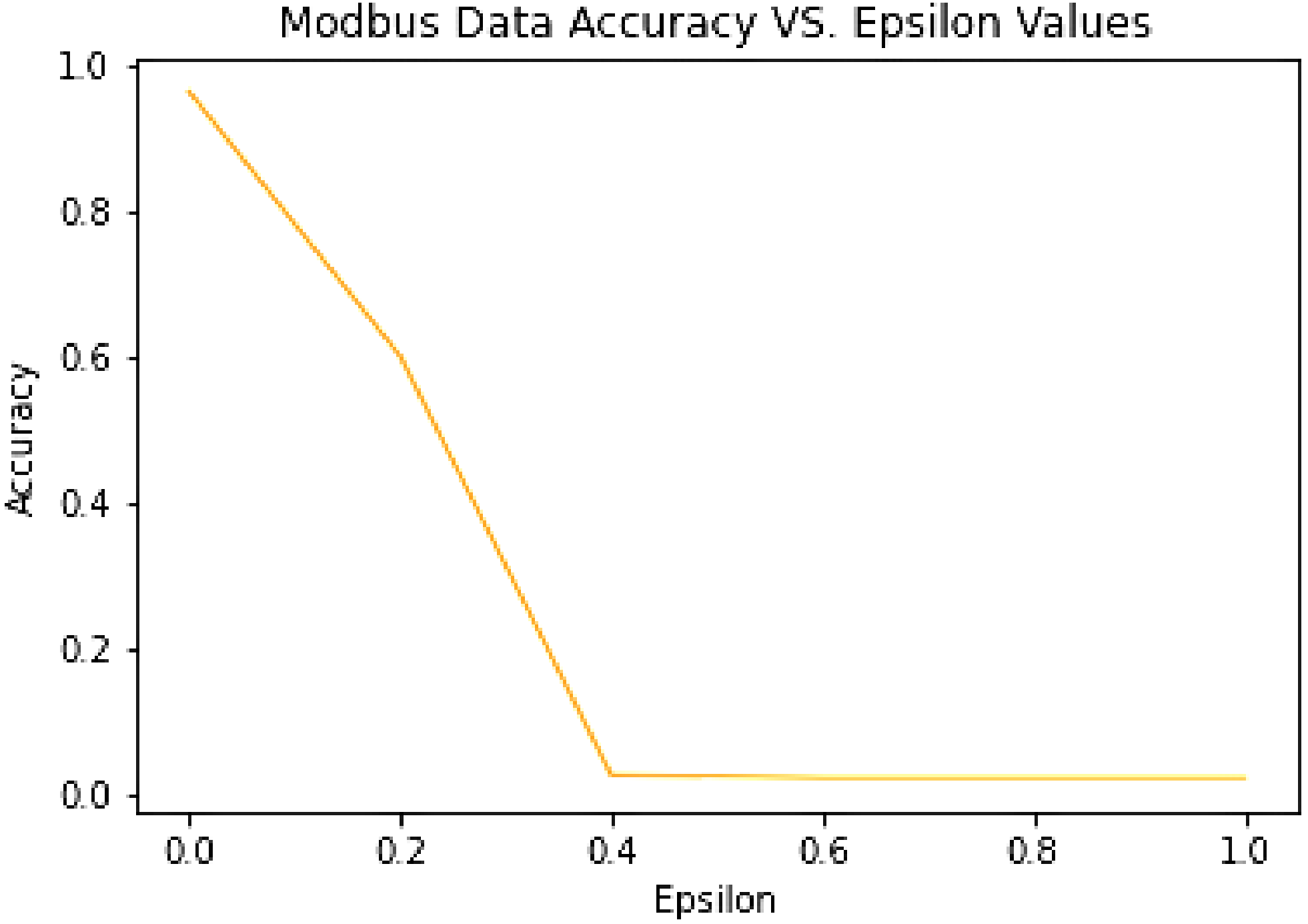}
    \caption{Impact of Epsilon values on the prediction accuracy (for Modbus dataset).}
    \label{fig:modbus-epsilon}
\end{figure}


These three models were subjected to adversarial attacks with different epsilon values (noise) ranging from 0 to 1. In order to assess the classification model’s robustness, this experiment used a common gradient-based attack called FGSM. The FGSM attack served as an initial attack to illustrate how this type of adversarial attack can gravely affect an outcome of a classification model. When all these models were tested for FGSM attack, the results clearly show how the adversarial attack has deteriorated the model performance. 

Tables~\ref{tab:results} and ~\ref{tab:matrics} show FGSM attacks with different epsilon values in Bot-IoT and Modbus datasets. The impact of epsilon values on the accuracy of our ANN classifier for all the three models is shown in Fig.~\ref{fig:epsilon-imbalanced-bot}, Fig.~\ref{fig:epsilon-balanced-bot} and Fig.~\ref{fig:modbus-epsilon}.  


\section{Defence strategies}
\label{Defence strategies}

We implement a popular defensive approach known as Adversarial training. Adversarial training is the approach of training an ML model with adversarial samples to increase the model's ability to detect adversarial attacks. An experiment conducted by \cite{goodfellow2014explaining} showed that adversarial training can minimise the impact of the attack when the data is perturbed by the adversary. Adversarial training can help strengthen the robustness of the model and can fend off future adversarial attacks. We use a defensive approach by retraining the model including adversarial samples in the train and test datasets in the form of percentages. These percentages act as the percentage of adversarial samples in the dataset. This defensive approach is related to our study because it demonstrates the model's ability to reduce the impact caused by the attack. With adversarial training, our model can produce better results than when it was trained with an unpolluted dataset. When the model was trained with more adversarial samples, it became robust and classified correctly. The percentage of adversarial samples is shown in Table~\ref{tab:my_label3}. From our experiments, we observe the rise in precision when the percentage of adversarial samples in an unpolluted dataset increases. We can see the results in Tables~\ref{table:exp_results_cm} and~\ref{table:exp_results}. These results depict that adding an adversarial sample into the training has proven to be an excellent defensive strategy against adversarial attacks. However, the train and test split are kept the same as in previous experiments, replacing the 10\%, and 20\% of the training samples with adversarial attacked samples, with the highest epsilon value (i.e., 1.0) for the FGSM attack. We have added 10\% and 20\% of adversarial samples while training the model. Adding more samples to the training data and using different adversarial attacks is a different direction of research. 


\begin{table}[h]
    \centering
    \caption{The composition of the data for defence experiments.}
    \begin{tabular}{l l l | l l}
    \toprule
        & & Total & Adversarial & Original \\
    \toprule
        Experiment 1 & Training Data & 70\% & 10\% & 90\% \\
        & Testing Data & 30\% & 100\% & 0\% \\
    \hline
        Experiment 2 & Training Data & 70\% & 20\% & 80\% \\
        & Testing Data & 30\% & 100\% & 0\% \\
    \bottomrule
    \end{tabular}
    \label{tab:my_label3}
\end{table}

\begin{table}[h]
    \centering
     \caption{Confusion matrices of the experiments.}
    \scalebox{1}{
    \begin{tabular}{l l l l l}
    \toprule
       Testing dataset name & TP & TN & FP & FN\\
    \toprule
        \multicolumn{5}{c}{Experiment 1 (10\% adversarial samples)} \\
    \toprule
        Bot-IoT & & & \\
        (imbalanced) & 0 & 880,335 & 0  &111 \\
    \hline
        Bot-IoT & & & \\
        (balanced) & 879,779 & 875,447 & 4,888  &555 \\
    \hline
        Modbus &5,216 &3,102 &1,064 &27 \\
    \toprule
        \multicolumn{5}{c}{Experiment 2 (20\% adversarial samples)} \\
    \toprule
        Bot-IoT & & & & \\
        (imbalanced) & 0 & 880,335 & 0 & 111 \\
    \hline
        Bot-IoT & & & & \\
        (balanced) & 879,642 & 875,367 & 4,968 & 692\\
    \hline
        Modbus & 5,216 & 3,102 & 1,064 & 27 \\
    \bottomrule
    \end{tabular}
    }
    \label{table:exp_results_cm}
\end{table}

\begin{table}[htb!]
    \centering
    \caption{Classification report of experiment with Bot-IoT and Modbus datasets.}
    \scalebox{1}{
    \begin{tabular}{l l l l}
    \toprule
       Testing dataset name& Precision  & Recall & F1-Score\\
       & (\%) & (\%)\\
    \toprule
        \multicolumn{4}{c}{Experiment 1 (10\% adversarial samples)}\\
    \toprule
        Bot-IoT & & & \\
        (imbalanced) & 0 & 0 & 0\\
    \hline
        Bot-IoT & & & \\
        (balanced) & 99.35 & 99.95 & 99.65\\
    \hline
        Modbus & 0.05 & 0.04 & 0.04\\
    \toprule
        \multicolumn{4}{c}{Experiment 2 (20\% adversarial samples)}\\
    \toprule
        Bot-IoT & & &\\
        (imbalanced) & 0 & 0 & 0\\
    \hline
        Bot-IoT & & &\\
        (balanced) & 99.44 & 99.92 & 99.68\\
    \hline
        Modbus & 83.06 & 99.49 & 90.53\\
    \bottomrule
    \end{tabular}
    }
    \label{table:exp_results}
\end{table}


\section{Conclusion and Future Work}
\label{conclusion}
Cyber Physical Systems (CPS) transformed the interaction with the physical world by connecting physical devices to the Internet and the cyber world. Deep learning-based intrusion detection has been widely used in such CPS environments to learn the baseline of normal behaviour and detect abnormal activities. However, advanced deep learning-based solutions are vulnerable to Adversarial Machine Learning (AML) attacks. In this paper, we addressed this issue and investigated the impact of AML attacks on intrusion detection for both IoT (Internet of Things) and large-scale Industrial IoT networks. For experimental purposes, two real-world datasets (i.e., Bot-IoT and Modbus) were used to evaluate the results. These datasets depicted actual industry scenarios where there are class imbalance scenarios. Therefore, we performed pre-processing of datasets to manage imbalanced data distribution. Fast Gradient Sign Method (FGSM) based adversarial attacks were studied in this paper, and the results showed that the performance of an Artificial Neural Networks (ANN) classifier is reduced under such FGSM attacks. Then, a defence strategy was proposed to improve the robustness of the deep learning by retraining it with adversarial samples. The results showed the improvement of deep learning robustness in FGSM samples. For future work in defensive strategies, adding adversarial samples from other types of adversarial attacks like Generative Adversarial Networks (GAN) would also be reviewed.

\section*{Acknowledgment}
The authors acknowledge the support of the Commonwealth of Australia and Cybersecurity Research Centre Limited.

\ifCLASSOPTIONcaptionsoff
  \newpage
\fi


\bibliographystyle{IEEEtran}
\bibliography{bare_jrnl}

\begin{thebibliography}{10}
\providecommand{\url}[1]{#1}
\csname url@samestyle\endcsname
\providecommand{\newblock}{\relax}
\providecommand{\bibinfo}[2]{#2}
\providecommand{\BIBentrySTDinterwordspacing}{\spaceskip=0pt\relax}
\providecommand{\BIBentryALTinterwordstretchfactor}{4}
\providecommand{\BIBentryALTinterwordspacing}{\spaceskip=\fontdimen2\font plus
\BIBentryALTinterwordstretchfactor\fontdimen3\font minus
  \fontdimen4\font\relax}
\providecommand{\BIBforeignlanguage}[2]{{%
\expandafter\ifx\csname l@#1\endcsname\relax
\typeout{** WARNING: IEEEtran.bst: No hyphenation pattern has been}%
\typeout{** loaded for the language `#1'. Using the pattern for}%
\typeout{** the default language instead.}%
\else
\language=\csname l@#1\endcsname
\fi
#2}}
\providecommand{\BIBdecl}{\relax}
\BIBdecl

\bibitem{pivoto2021cyber}
D.~Pivoto, L.~de~Almeida, R.~Righi, J.~Rodrigues, A.~Lugli,
  and A.~Alberti, ``Cyber-physical systems architectures for industrial
  internet of things applications in industry 4.0: A literature review,''
  \emph{Journal of manufacturing systems}, vol.~58, pp. 176--192, 2021.

\bibitem{yaacoub2020cyber}
J.~Yaacoub, O.~Salman, H.~Noura, N.~Kaaniche, A.~Chehab, and M.~Malli,
  ``Cyber-physical systems security: Limitations, issues and future trends,''
  \emph{Microprocessors and microsystems}, vol.~77, 2020.

\bibitem{sharma2022role}
A.~Sharma, V.~Burman, and S.~Aggarwal, ``Role of iot in industry 4.0,'' in
  \emph{Advances in Energy Technology}, Springer, pp. 517--528, 2022.

\bibitem{pal2017design}
S.~Pal, M.~Hitchens, and V.~Varadharajan, ``On the design of security
  mechanisms for the internet of things,'' in \emph{Eleventh International
  Conference on Sensing Technology (ICST)}, IEEE, pp. 1--6, 2017.

\bibitem{pal2021analysis}
S.~Pal and Z.~Jadidi, ``Analysis of security issues and countermeasures for the
  industrial internet of things,'' \emph{App. Sciences}, vol.~11, no.~20, 2021.

\bibitem{da2019internet}
K.~Costa, J.~Papa, C.~Lisboa, R.~Munoz, and V.~Albuquerque, ``Internet of things: A survey on machine learning-based
  intrusion detection approaches,'' \emph{Computer Networks}, vol. 151, pp. 147--157, 2019.

\bibitem{al2020survey}
M.~Garadi, A.~Mohamed, A.~Ali, X.~Du, I.~Ali, and M.~Guizani, ``A
  survey of machine and deep learning methods for internet of things (iot)
  security,'' \emph{IEEE Communications S\&T}, vol.~22, pp. 1646--1685, 2020.

\bibitem{haylett2021system}
G.~Haylett, Z.~Jadidi, and K.~Thanh, ``System-wide anomaly detection of
  industrial control systems via deep learning and correlation analysis,'' in
  \emph{AI Applications and Innovations}, Springer, pp. 362--373, 2021.

\bibitem{anthi2021hardening}
E.~Anthi, L.~Williams, A.~Javed, and P.~Burnap, ``Hardening machine learning
  denial of service (dos) defences against adversarial attacks in iot smart
  home networks,'' \emph{Computers \& Security}, 2021.

\bibitem{ozbulak2020perturbation}
U.~Ozbulak, M.~Gasparyan, W.~Neve, and A.~Messem, ``Perturbation
  analysis of gradient-based adversarial attacks,'' \emph{Pattern Recognition
  Letters}, vol. 135, pp. 313--320, 2020.

\bibitem{goodfellow2014explaining}
I.~Goodfellow, J.~Shlens, and C.~Szegedy, ``Explaining and harnessing
  adversarial examples,'' \emph{arXiv Preprint}, 2014.

\bibitem{papernot2016limitations}
N.~Papernot, P.~McDaniel, S.~Jha, M.~Fredrikson, Z.~Celik, and A.~Swami,
  ``The limitations of deep learning in adversarial settings,'' in \emph{Euro S\&P},
  IEEE, pp. 372--387, 2016.

\bibitem{moosavi2016deepfool}
S.~Dezfooli, A.~Fawzi, and P.~Frossard, ``Deepfool: a simple and
  accurate method to fool deep neural networks,'' in \emph{computer vision and pattern recognition conference}, IEEE, pp.
  2574--2582, 2016.

\bibitem{carlini2017towards}
N.~Carlini and D.~Wagner, ``Towards evaluating the robustness of neural
  networks,'' in \emph{IEEE symposium on security and privacy (SP)}, IEEE, pp. 39--57, 2017.

\bibitem{abiodun2018state}
O.~Abiodun, A.~Jantan, A.~Omolara, K.~Dada, N.~Mohamed, and
  H.~Arshad, ``State-of-the-art in artificial neural network applications: A
  survey,'' \emph{Heliyon}, vol.~4, no.~11, 2018.

\bibitem{wu2020anomaly}
D.~Wu, et~al., ``Anomaly detection based on rbm-lstm neural network for cps in
  advanced driver assistance system,'' \emph{ACM Transactions on Cyber-Physical
  Systems}, vol.~4, no.~3, pp. 1--17, 2020.

\bibitem{luo2021deep}
Y.~Luo, Y.~Xiao, L.~Cheng, G.~Peng, and D.~Yao, ``Deep learning-based anomaly
  detection in cyber-physical systems: Progress and opportunities,'' \emph{ACM
  Computing Surveys (CSUR)}, vol.~54, no.~5, pp. 1--36, 2021.

\bibitem{kim2021survey}
S.~Kim and K.~Park, ``A survey on machine-learning based security design for
  cyber-physical systems,'' \emph{App. Sci.}, vol.~11, no.~12, 2021.

\bibitem{mujeeb2021machine}
C.~Ahmed, M.~Umer, B.~Binte~Liyakkathali, M.~Jilani, and
  J.~Zhou, ``Machine learning for cps security: applications, challenges and
  recommendations,'' in \emph{Machine Intelligence and Big Data Analytics for
  Cybersecurity Applications}, Springer, pp. 397--421, 2021.

\bibitem{jadidi2020securing}
Z.~Jadidi, A.~Dorri, R.~Jurdak, and C.~Fidge, ``Securing manufacturing using
  blockchain,'' in \emph{International Conference on Trust,
  Security and Privacy in Computing and Communications (TrustCom)}, IEEE, pp. 1920--1925, 2020.

\bibitem{zhang2021label}
H.~Zhang, N.~Cheng, Y.~Zhang, and Z.~Li, ``Label flipping attacks against naive
  bayes on spam filtering systems,'' \emph{App. Intelli.}, pp. 1--12,
  2021.

\bibitem{patil2021improving}
S.~Patil, V.~Varadarajan, D.~Walimbe, S.~Gulechha, S.~Shenoy, A.~Raina, and
  K.~Kotecha, ``Improving the robustness of ai-based malware detection using
  adversarial machine learning,'' \emph{Algo.}, vol.~14, no.~10, 2021.

\bibitem{chauhan2020polymorphic}
R.~Chauhan and S.~Heydari, ``Polymorphic adversarial ddos attack on ids
  using gan,'' in \emph{International Symposium on Networks, Computers and
  Communications (ISNCC)}, IEEE, pp. 1--6, 2020.

\bibitem{song2020fda}
Y.~Song, T.~Liu, T.~Wei, X.~Wang, Z.~Tao, and M.~Chen, ``Fda: Federated defense
  against adversarial attacks for cloud-based iiot applications,'' \emph{IEEE
  Trans. on Industrial Infor.}, vol.~17, no.~11, pp. 7830--7838,
  2020.

\bibitem{dankwa2021securing}
S.~Dankwa and L.~Yang, ``Securing iot devices: A robust and efficient deep
  learning with a mixed batch adversarial generation process for captcha
  security verification,'' \emph{Electronics}, vol.~10, no.~15, 2021.

\bibitem{qiu2020adversarial}
H.~Qiu, T.~Dong, T.~Zhang, J.~Lu, G.~Memmi, and M.~Qiu, ``Adversarial attacks
  against network intrusion detection in iot systems,'' \emph{IEEE Internet of
  Things Journal}, vol.~8, no.~13, pp. 10327--10335, 2020.

\bibitem{mode2020crafting}
G.~Mode and K.~Hoque, ``Crafting adversarial examples for deep learning
  based prognostics (extended version),'' \emph{arXiv Preprint}, 2020.

\bibitem{yang2021iot}
Z.~Yang, I.~Abbasi, F.~Algarni, S.~Ali, and M.~Zhang, ``An iot time series
  data security model for adversarial attack based on thermometer encoding,''
  \emph{Security and Communication Networks}, 2021.

\bibitem{luo2020adversarial}
Z.~Luo, S.~Zhao, Z.~Lu, Y.~Sagduyu, and J.~Xu, ``Adversarial machine
  learning based partial-model attack in iot,'' in \emph{2nd
  ACM Workshop on Wireless Security and Machine Learning}, pp. 13--18, 2020. 

\bibitem{ibitoye2019analyzing}
O.~Ibitoye, O.~Shafiq, and A.~Matrawy, ``Analyzing adversarial attacks against
  deep learning for intrusion detection in iot networks,'' in \emph{GlobeCom}, IEEE, 2019, pp. 1--6.

\bibitem{wang2018deep}
Z.~Wang, ``Deep learning-based intrusion detection with adversaries,''
  \emph{IEEE Access}, vol.~6, pp. 38367--38384, 2018.

\bibitem{dataset-link}
\BIBentryALTinterwordspacing
UNSW-Canberra-Australia, ``The ton\_iot datasets.'' [Online]. Available:
  \url{https://research.unsw.edu.au/projects/toniot-datasets (January 2022)}
\BIBentrySTDinterwordspacing

\bibitem{koroniotis2019towards}
N.~Koroniotis, N.~Moustafa, E.~Sitnikova, and B.~Turnbull, ``Towards the
  development of realistic botnet dataset in the internet of things for network
  forensic analytics: Bot-iot dataset,'' \emph{Future Generation Computer
  Systems}, vol. 100, pp. 779--796, 2019.

\bibitem{alsaedi2020ton_iot}
A.~Alsaedi, N.~Moustafa, Z.~Tari, A.~Mahmood, and A.~Anwar, ``Ton\_iot
  telemetry dataset: A new generation dataset of iot and iiot for data-driven
  intrusion detection systems,'' \emph{IEEE Access}, vol.~8, 2020.

\bibitem{abdulwahab2022feature}
H.~Abdulwahab, S.~Ajitha, and M.~ASaif, ``Feature selection techniques
  in the context of big data: taxonomy and analysis,'' \emph{Applied
  Intelligence}, pp. 1--46, 2022.

\bibitem{pears2014synthetic}
R.~Pears, J.~Finlay, and A.~Connor, ``Synthetic minority over-sampling
  technique (smote) for predicting software build outcomes,'' \emph{arXiv
  Preprint}, 2014.

\bibitem{papadopoulos2021launching}
P.~Papadopoulos, O.~Essen, N.~Pitropakis, C.~Chrysoulas,
  A.~Mylonas, and W.~Buchanan, ``Launching adversarial attacks against
  network intrusion detection systems for iot,'' \emph{Journal of Cybersecurity
  and Privacy}, vol.~1, no.~2, pp. 252--273, 2021.

\bibitem{mahfuz2021mitigating}
R.~Mahfuz, R.~Sahay, and A.~Gamal, ``Mitigating gradient-based adversarial
  attacks via denoising and compression,'' \emph{arXiv Preprint}, 2021.

\bibitem{papernot2016technical}
N.~Papernot, et~al., ``Technical report on the
  cleverhans v2. 1.0 adversarial examples library,'' \emph{arXiv Preprint}, 2016.

\bibitem{git-code}
\BIBentryALTinterwordspacing
``Github-code.'' [Online]. 
  \url{github.com/NitheshNayak/AnomalyDetection\\CyberPhysicalSystems.git}
\BIBentrySTDinterwordspacing

\end{thebibliography}

\end{document}